\documentclass[runningheads]{llncs}

\usepackage{cite}
\usepackage{amsmath,amssymb,amsfonts}
\usepackage{graphicx}
\usepackage[caption=false]{subfig}

\usepackage[outdir=./]{epstopdf}
\usepackage{textcomp}
\usepackage{xcolor}
\usepackage{url}

\usepackage{enumitem}

\usepackage{algorithm}
\graphicspath{ {images/} }
\usepackage{algorithmicx}
\usepackage[noend]{algpseudocode}

\usepackage[utf8]{inputenc}
\usepackage[english]{babel}
\usepackage{lipsum}

\begin{document}
\title{IaaS Signature Change Detection with Performance Noise}

%\subtitle{resubmission of paper \#35 \vspace{-15mm}}

\author{ Sheik Mohammad Mostakim Fattah \and
Athman Bouguettaya}
\authorrunning{S. Fattah et al.}
\institute{School of Computer Science, University of Sydney, Sydney, Australia
\email{\{sheik.fattah,athman.bouguettaya\}@sydney.edu.au}}

\maketitle              % typeset the header of the contribution

\begin{abstract}

We propose a novel framework to detect changes in the performance behavior of an IaaS service. The proposed framework leverages the concept of the IaaS signature to represent an IaaS service's long-term performance behavior. A new type of performance signature called categorical IaaS signature is introduced to represent the performance behavior more accurately. A novel performance noise model is proposed to accurately identify IaaS performance noise and accurate changes in the performance behavior of an IaaS service. A set of experiments based on real-world datasets is carried out to evaluate the effectiveness of the proposed framework. 


\end{abstract}

\section{Introduction}

Infrastructure-as-a-Service (IaaS) models offer various computational resources such as CPU, memory, storage, and network are offered as Virtual Machines (VMs) \cite{chaisiri2012optimization}. Large organizations tend to utilize IaaS cloud services on a \textit{long-term basis} (e.g., 1 - 3 years). Most leading IaaS cloud providers such as Amazon, Google, and Microsoft offer significant discounts on long-term subscriptions. Selecting a service for a long-term period is a key decision for many consumers. Committing to a service for a long-term period that may perform poorly, may cause loss of revenue. Therefore, it is important for a consumer to know the performance of an IaaS service.

IaaS providers typically reveal \textit{limited} performance information in their advertisements due to \textit{market competition} and \textit{business secrecy} \cite{wenmin2011history}. For example, most IaaS advertisements do not contain actual vCPU (virtual CPU) speed, memory bandwidth, or VM startup time information. The performance of a VM may change over time due to the dynamic nature of the cloud \cite{iosup2011performance}. Therefore, advertised performance information may not reflect the true service performance for a certain time.

An effective way to deal with the limited performance information is to leverage free trials \cite{wang2018testing}. Most IaaS providers promote \textit{free short-term trials} and invite potential consumers to test their application in the cloud. Therefore, a consumer may run its application workload on different IaaS cloud services and compare their performance. free trial experiences, however, do not provide sufficient information to make a long-term commitment \cite{fattah2020icws}. The performance of IaaS services changes \textit{periodically} due to the multi-tenant nature of the cloud \cite{iosup2011performance}. The observed performance in a trial in one month may change if the trial is performed in a different month. Therefore, making a long-term commitment based on only short trials may lead to a poor selection.

\textit{IaaS performance signatures} provide an effective alternative to deal with the unknown service performance variability for the long-term selection \cite{mi2008analysis,fattah2020icws}. The performance signature of an IaaS service represents its \textit{expected} performance behavior over a long period of time. For instance, a signature of a VM may indicate that its response time is expected to increase by 10\% in January than the response time in December. A consumer's trial experience of a service and its corresponding signature can be utilized together to make a better selection for the long-term period. A signature-based IaaS selection approach is proposed that generates IaaS signatures using the experience of past trial users over different periods of a year \cite{fattah2020icws}. However, most existing selection approaches do not consider the long-term changes in IaaS performance behavior where the signature may need to be re-evaluated periodically. \textit{The focus of this work is to detect changes in long-term IaaS performance behavior.}

An IaaS service's performance behavior may change over time due to a number of reasons \cite{mi2008analysis,chaki2020fine}. For instance, a provider may upgrade its infrastructure or change its multi-tenant management policy resulting in the change of service performance \cite{leitner2016patterns}. Therefore, detecting the change of IaaS performance is important to ensure that its signature reflects the \textit{current} performance behavior of the service. \textit{We focus on the detection of changes in IaaS performance behavior as represented by its signature}. In this case, the IaaS performance signature may need to be updated to be representative of the new performance profile of the service. 

There are two key challenges in IaaS performance Signature change detection. The first challenge is detecting the point in time where the signature needs to be re-evaluated. A change may occur at any point in time. Therefore, it is required to identify change points in time where there is a high probability of performance change occurrence. This is typically known as the \textit{Change Point Detection} problem \cite{aminikhanghahi2017survey}. The second challenge is to differentiate between the \textit{noise} and \textit{true changes} in performance. Noise typically indicates the irregular or anomalous behavior in service performance that may not be the long-term performance changes \cite{moens2019learning}. For instance, a major power failure may impact the service performance at a time without necessarily indicating a long-term performance change. Noise in IaaS performance is very common due to the dynamic nature of the cloud. 

\textit{To the best of our knowledge, existing research has not given enough attention to the long-term IaaS performance change detection problem \cite{fattah2020icws}}. An IaaS performance change detection framework is proposed  that utilizes an ECA model to detect changes in IaaS performance \cite{fattah2020event}. However, it does not consider noise in IaaS performance during the change detection. Therefore, \textit{the focus of this paper is to distinguish the true changes in IaaS performance from the changes that are caused by performance noise}.

Noise in signal processing generally represents the unwanted disturbance in
electrical signals, which is usually generated during the capture, storage, trans-
mission, processing, or conversion of the signal. In the case of IaaS cloud, noise can be generated from co-tenants, system upgrade, or temporary service disruptions \cite{varadarajan2012resource}. \textit{We propose a novel framework to detect changes in IaaS performance signature by accurately detecting noise and true changes in IaaS performance.} The proposed framework introduces a new type of IaaS performance signature called \textit{categorical IaaS signature}. The categorical IaaS signature models performance behavior more \textit{accurately} than the \textit{general IaaS signature} introduced in \cite{fattah2020icws} as the general IaaS signature does not consider the effect of different categories of workloads, i.e., CPU-intensive, I/O-intensive, and memory-intensive on IaaS performance. The proposed framework utilizes a heuristic-based approach to determine noise in IaaS performance. In this approach, the categorical signature and the general signature are utilized to define performance noise bandwidth. The performance noise bandwidth is updated over time to detect performance changes more accurately. The key contributions are summarized as follows:

\begin{itemize} 
    \item A new type of IaaS performance signature called Categorical IaaS Signature that models an IaaS service's long-term performance behavior based on different categories of workloads.
    
    \item A novel performance noise model that defines the noise bandwidth based on the categorical and general IaaS signatures. 
    
    \item A performance change detection model that leverages the proposed performance noise model to detect changes in IaaS performance. 
\end{itemize} 

\section{IaaS Performance Signatures}

We overview the general and categorical IaaS performance signatures, their representations, and generation techniques.

\subsection{General IaaS Performance Signatures}

The general IaaS performance signature is first introduced in \cite{fattah2020icws}. The general signature of an IaaS service is represented based on its \textit{relative} performance changes over time, i.e., how much a service's performance may increase or decrease in one time compared to another time. For example, the general signature of a VM may inform that its response time is expected to increase by 5\% on weekend nights than regular weekdays. The general signature mainly focuses on the effect of seasonality on IaaS performance. It assumes that the effect of different types of workload on the observed performance is not substantial compared to the effect of seasonal performance variability. Therefore, this signature is called general signature as it considers all types of workloads equally. Note that the signature does not tell the exact performance of a service. Therefore, a consumer is unable to select a service based on only its signature. Instead, the consumer needs to perform the trial with its application workloads and utilize the trial experience and the IaaS signature to estimate the long-term service performance \cite{fattah2020icws}. 

\begin{definition} {General IaaS Performance Signature:} An IaaS performance signature is a temporal representation of relative performance changes of an IaaS service over a long period.

\end{definition}

The general IaaS performance signature is represented by a set of QoS parameters that are relevant to the service. The \textit{relevant} QoS attributes are defined by the most important QoS attributes to measure the performance of a particular type of IaaS service \cite{fattah2020icws}. For example, data read/write throughput and disk latency are the key QoS attributes for virtual storage services.

We denote the general signature of a service as $S=\{S_1, S_2,...S_n\}$, where $n$ is the number of QoS attributes in the signature. Each $S_i$ corresponds to a QoS attribute. Each $S_i$ denotes a time series for $t$ period which is represented as $S_i = \{s_{i1},s_{i2},......s_{it}\}$. Here, $s_it$ is the relative performance of the provider at the time $t$ for a particular QoS attribute. We use the following representation to denote a signature:

\scriptsize
\begin{gather}
%\vspace{-.5cm}
 S =
  \begin{bmatrix}
   s_{11} & s_{12} & .. & s_{1t} \\
   s_{21} & s_{22} & .. & s_{2t}  \\
  s_{31} & s_{13} & .. & s_{3t}  \\
   .. & .. & ... \\
   s_{n1} & s_{n2} & .. & s_{nt}  \\
   \end{bmatrix}
   \label{eqn:signature}
   %\vspace{-.2cm}
\end{gather}

\normalsize where each row corresponds to the QoS signature of $S_i$ and each column represents a timestamp $t$. From the equation \ref{eqn:signature}, we see that a signature may include several QoS attributes. However, we describe the proposed approach using only one QoS attribute in this work, i.e., throughput of an IaaS service for simplicity. However, the proposed approach is applicable for more than one QoS attribute of IaaS performance signatures.

\subsection{General IaaS Performance Signature Generation}

\label{sec:sig_gen}

\begin{figure} [t]
%\vspace{-.4cm}
    \centering
    \includegraphics[width=.8\textwidth]{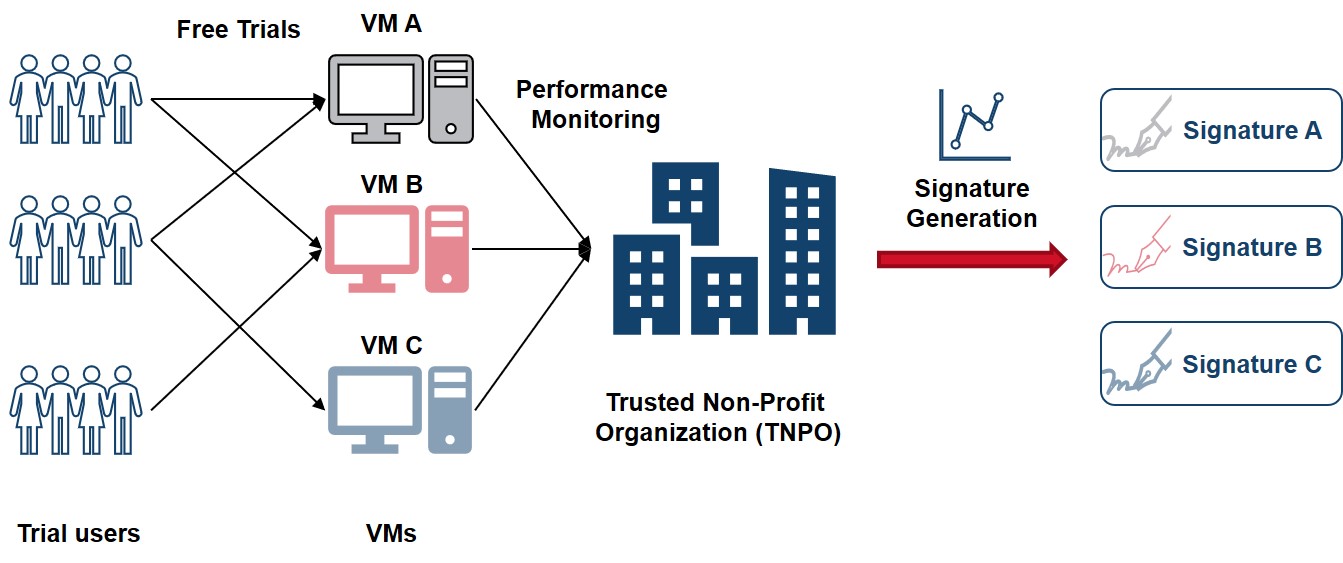}
    \caption{IaaS performance signature generation}
    \label{fig:tnpo}
    \vspace{-.5cm}
\end{figure}

\label{sec:sig}

It is important to note that, the past trial users may not want to share their experience publicly to protect their privacy, security, and the conflict of interests with the provider \cite{zhu2015privacy}. However, they may share their trial experience with a \textit{Trusted} Non-Profit Organization (TNPO) for a limited period to help new consumers in the selection \cite{van2012trusted}. Examples of such TNPOs are available in public sectors where privacy-sensitive information about individuals needs to be shared to deliver better services. For instance, health research institutes often collect data about individual patients to improve health services. TNPOs are responsible for data \textit{integration} and \textit{distribution} of collective knowledge without revealing individual's privacy-sensitive information.

\textit{We assume that the past trial users who have utilized some IaaS services share their experience with a TNPO for a limited period of time.} The TNPO generates IaaS performance signatures based on the aggregated experience of past trial users and deletes the users' data afterward. Let us assume that there are three IaaS providers ($A$, $B$, and $C$) who offer three VMs ($VM_a$, $VM_b$, and $VM_c$) with similar configurations (e.g., resource capacity, location) for free short-term trials as shown in Fig. \ref{fig:tnpo}. There are past users who utilized the VMs to find the performance over different periods of time. The trial users do not want to share their trial experience publicly. However, each trial user shares its experience with a TNPO for a short period. The TNPO generates the signature to identify the long-term performance variability of each VM. The TNPO has to delete users' experience once the signatures are computed. A signature provides an aggregated view of a VM's long-term performance variability. It is not possible to derive individual trial experience from the signature. As a result, \textit{the TNPO does not violate the privacy of past trial users}.

We create IaaS performance signatures in a way that requires less detailed performance information about the service performance and the past trial users and yet useful enough to make a long-term selection. Let us assume that $k$ number of past trial users share their observed trial performance $Q_{k}$ over the period $T$ for a service. Here, $Q_{k}$ refers to the performance observed by the $k$th consumer for the QoS attribute $Q$ over the period $T$. We denote $Q_{k}$ as $Q_{k}=\{q_{1k}, q_{2k},.., q_{tk}\}$. The following steps are performed to generate the signature for the QoS attribute $Q$:

\begin{enumerate}[itemsep=0ex, leftmargin=*]
    \item For a QoS attribute $Q$, the performance observed by the trial users is collected over time $T$.
    \item At each timestamp $t \in T$, the average performance observed by $k$ number of consumers is measured for $Q$. The average performance is denoted by $\overline{Q_{k}}$.
 
\end{enumerate}

The value of $s_nt$ at any $t$ represents the average QoS performance compare to any other time $t'$ in Equation \ref{eqn:signature}. This representation of the signature offers two benefits. First, the use of signature becomes easier once a consumer has utilized free trials based on its workloads. The performance for any other time can be found by comparing the ratio between the trial month and other times. Second, signatures can be stored and updated easily over time as it does not require storing detailed information about consumers' trial. 

\subsection{Categorical IaaS Performance Signatures}

In this subsection, we introduce a new type of signature called categorical IaaS performance signature. For simplicity, we refer to the categorical IaaS performance signature as the categorical signature and the general IaaS performance signature as the general signature. The motivation behind creating the categorical signature is to produce a more accurate signature that captures the effect of different types of workloads on IaaS performance behavior. The performance of an IaaS service may depend on the workload it runs \cite{feitelson2002workload}. Therefore, IaaS providers often advertise CPU-intensive, memory-intensive, or network-intensive VMs. For instance, Amazon EC2 offers a wide range of compute-optimized, storage-optimized, and memory-optimized instances.

IaaS workloads can be categorized based on several workload parameters such as resource requirements, request arrival rates, and workload distribution. Without loss of generality, we only consider resource requirements as workload parameters for categorization in this work. Therefore, workload categories will be CPU-intensive, memory-intensive, and I/O intensive. The proposed workload categorization is applicable for any other workload parameters. Let us assume there are $N_c$ types of workload based on resource requirements of consumer requests. Therefore, we create $N_c$ number of categorical signatures. A categorical signature is  represented as:

\scriptsize
\begin{gather}
S_c =
  \begin{bmatrix}
   s_{11} & s_{12} & .. & s_{1t} \\
   s_{21} & s_{22} & .. & s_{2t}  \\
  s_{31} & s_{13} & .. & s_{3t}  \\
   .. & .. & ... \\
   s_{n1} & s_{n2} & .. & s_{nt}  \\
   \end{bmatrix}
   \label{eqn:cat_signature}
   %\vspace{-.2cm}
\end{gather}
\normalsize

where $S_c$ represents the signature for $c$ categories of workloads. Here, $c$ is one of the categories in $N_c$. Rest of the attributes of Equation \ref{eqn:cat_signature} are same as the general signature in Equation \ref{eqn:signature}.

\subsection{Categorical IaaS Performance Signature Generation}

The key difference between the categorical signature generation and the general signature generation is the consideration of different workload categories. First, we define a set of categories ($C$) based on the resource requirements where $C = \{1, 2, 3, ... N_c\}$. For each category, we define the criteria that determine the category of each request (workload). Let us assume that a consumer's request has $R$ number of attributes where each attribute denotes a resource in the VM such as vCPU, storage, or memory. For each attribute ($a$), we define a minimum resource requirement $M_a$. If a request has more than $M_a$ amount of resource requirement for the attribute $a$, we consider that request as $a$-intensive request. For example, if a request has 80\% of CPU usage requests, then we consider that request as a CPU-intensive request. According to this approach, a request can be in multiple categories of workloads. The minimum resource requirement for each attribute is defined experimentally by the TNPO for each cloud provider, i.e., the different threshold is considered as the minimum resource requirement for each category to find the most effective threshold. Once we define the category for each workload, we create the categorical signature as follows:    

\begin{enumerate}[itemsep=0ex, leftmargin=*]
    \item For a QoS attribute $Q$, the performance observed by the trial users is collected over time $T$.
    \item For each category $a$ at each trial length $\delta T$, we identify $k$ number of $a$-intensive requests. The average performance ($\overline{Q_{k}}$) is measured for each QoS attribute. 

\end{enumerate}

We computed the average performance of a QoS attribute to obtain the IaaS signature. The signature should reflect performance behavior of the service for all types of workloads. However, the performance of a service may depend on its workload. Therefore, we introduced the categorical signature to represent signature for similar categories of workloads.  It is not practical to define signature for every workload. Therefore, we utilized the average performance as it is a good approximation of the performance behavior. We improve the accuracy of the signature by adjusting the noise bandwidth over time.

\section{Proposed Change Detection Framework}

\begin{figure} [t]
%\vspace{-.6cm}
    \centering
    \includegraphics[width=.8\textwidth]{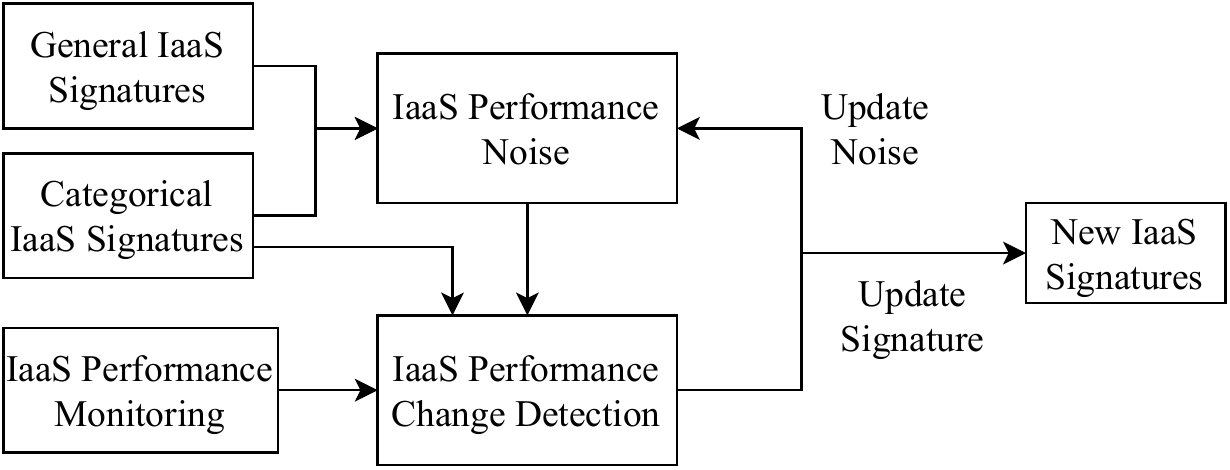}
    \caption{IaaS performance change detection framework}
    \label{fig:frame}
    \vspace{-.6cm}
\end{figure}

In this section, we discuss the proposed change detection framework as shown in Fig. \ref{fig:frame}. The proposed framework consists of two key components: a) IaaS performance noise and b) IaaS performance change detection. The performance noise is initially defined by the general signature and the categorical signature. The performance noise is then updated dynamically based on the observed performance of the free trial users. The change detection framework utilizes the knowledge of IaaS performance noise and the categorical IaaS signature to detect changes in the categorical signature based on the observed performance by the free trial users. The change detection framework updates the knowledge about the performance noise based on the observed performance over time.

\subsection{IaaS Performance Noise}

A key step in identifying changes in IaaS performance is to accurately determine the noise in IaaS performance. We define the noise in IaaS performance as the deviation from the expected performance behavior as represented by the signature of an IaaS service. The key challenge in defining the performance noise is to determine the amount of performance fluctuation from the expected performance behavior. A boundary must be defined, which will determine whether the observed performance fluctuations can be considered as the noise or a permanent change in the performance behavior. In signal processing, image processing, and other domains, there are many approaches to define and detect different types of noises such as White noise, Gaussian noise, and Salt and pepper noise. \textit{To the best of our knowledge, there is no definitive way of defining noise in the case of IaaS performance behavior.} Therefore, we propose a heuristic-based approach using the general signature and the categorical signature to define the initial performance noise boundary of an IaaS service. We call it \textit{IaaS performance noise bandwidth}. The noise bandwidth is updated over time based on the observed performance behavior of an IaaS service.  The performance noise bandwidth is defined as follows:

\begin{definition} {IaaS Performance Noise Bandwidth:} The surrounding area created by the acceptable fluctuation from the expected performance of an IaaS service is the IaaS performance noise bandwidth of that service. 

\end{definition}

\begin{figure} [t]
%\vspace{-.35cm}
    \centering
    \includegraphics[width=.6\textwidth]{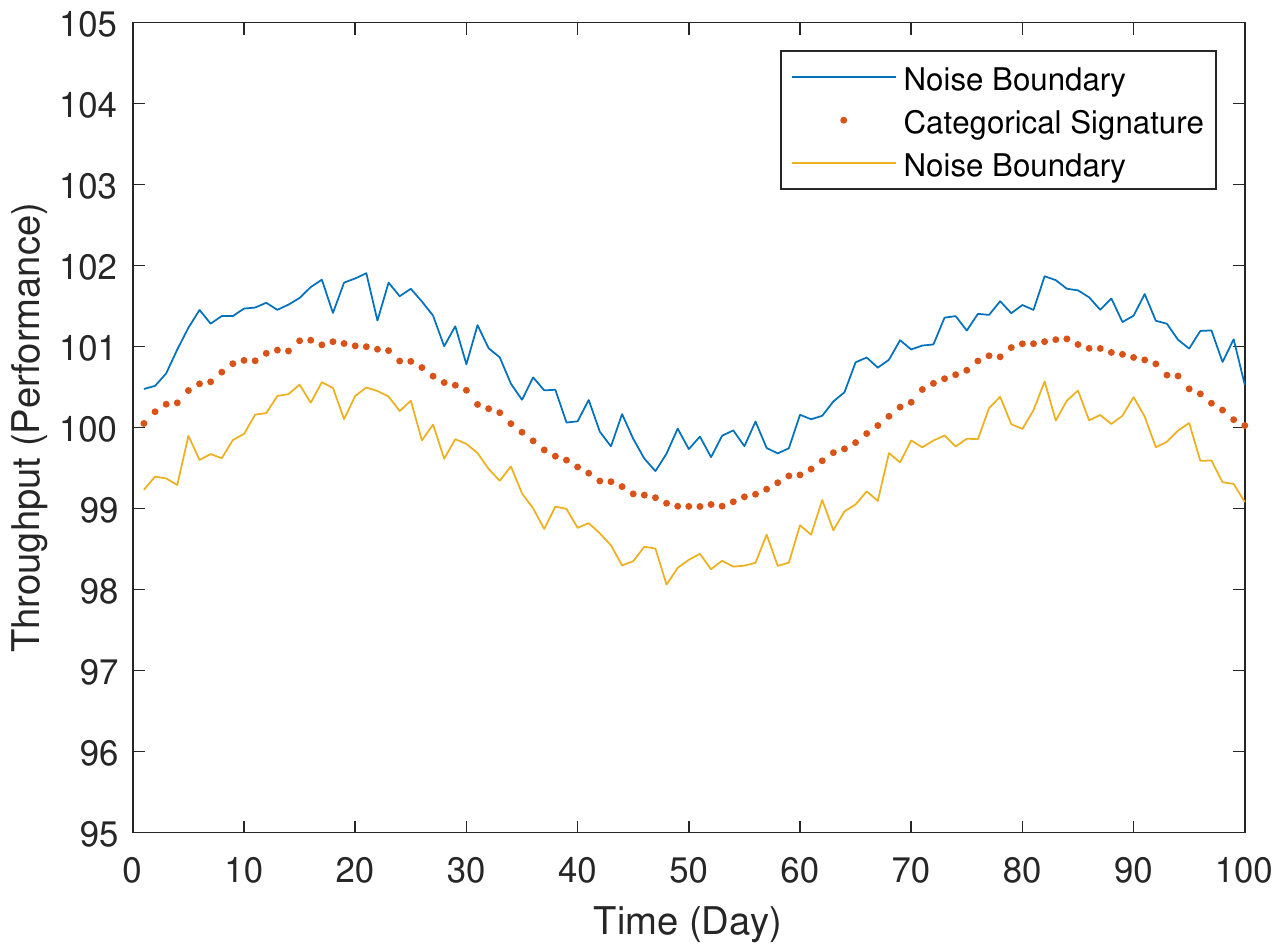}
    \caption{IaaS performance noise bandwidth}
    \label{fig:noise}
    %\vspace{-.6cm}
\end{figure}

The amount of acceptable fluctuation is initially defined by the general signature and the categorical signature as shown in Fig. \ref{fig:noise}. The distance between the general signature and the categorical signature $D$ is computed for each timestamp by the following equation:

\begin{equation}
    D = dist(S,S_c) =  \forall(S_i, S_{ci}) \; abs(S_i, S_{ci})
\end{equation}

where $S$ is the general signature, $S_c$ is the categorical signature, $S_i$ is the value of the general signature at $i$th timestamp, and $S_{ci}$ is the value of the categorical signature at $i$th timestamp. The dist function is computed based on the absolute distance between $S_i$ and $S_{ci}$. $D$ is then considered as the acceptable deviation from the expected performance as represented by the categorical signature. Therefore, any observed performance that has the maximum deviation $D$ from the categorical signature is considered noisy performance. The data for Fig. \ref{fig:noise} are obtained synthetically to demonstrate the performance noise.

\subsection{IaaS Performance  Change Detection}

Detecting changes in performance requires monitoring the current performance behavior of an IaaS service. We assume that the TNPO continues to monitor the experience of free trial users after creating the signatures. When most of the users' experience does not \textit{match} with the corresponding categorical signature, the existing signature needs to be re-computed. We represent the signatures and the trial experience as time series. Therefore, the matching of trial experience and signature has two parts: a) distance and b) shape. The distance $D'$ is computed based on the absolute distance between the categorical signature and the trial experience for a given trial period $T$ using the following equation: 

\begin{equation}
   D' = \forall_{i\in T} \; abs(S_{ci}, E_i) 
\end{equation}

where $S_{ci}$ and $E_i$ are the value of the categorical signature and observed performance at timestamp $i$. We utilize the pearson correlation coefficient to measure the shape based similarity using the following equation:

\begin{equation}
    S(E, S_c)^{PCC} = \frac{\sum_{i=1}^T (S_{ci} - \Bar{S}) (E_i - \Bar{E}) }{\sqrt{ (S_{ci} - \Bar{S_c})^2 } \sqrt{(E_i - \Bar{E})^2}}
\end{equation}

where $\Bar{E}$ and $\Bar{S_c}$ are the average of $E$ and $S_c$ in period $T$. When the observed performance of a user has a distance from the categorical signature within the performance noise bandwidth, and the shape of the observed performance is similar to the categorical signature, we assume that there is no change in performance. We identify the following cases during the matching based on the shape and the distance:

\begin{enumerate}
    \item Case 1: Most of the users' observed performance is within the noise bandwidth, and the shape of the performance is similar to the corresponding categorical signatures. In this case, no action is taken.
    \item Case 2: Most of the users' observed performance is outside the noise bandwidth, and the shape of the performance is not similar to the corresponding categorical signatures. In this case, signatures are required to be recomputed.  
    \item Case 3: Most of the users' observed performance is within the noise bandwidth, and the shape of the performance is not similar to the corresponding signatures. In this case, we reduce the size of the performance noise bandwidth. 
    \item Case 4: Most of the users' observed performance is outside but adjacent to the noise bandwidth, and the performance shape is similar to the corresponding categorical signatures. In this case, we increase the size of the performance noise bandwidth. 
\end{enumerate}

Let us assume that the noise bandwidth at timestamp $t$ is defined by $d^+$ and $d^-$ where $d^+$ is the distance from the categorical signature to the noise boundary on the upper side of the Y-axis, and $d^-$ is the distance from the categorical signature to the noise boundary on the downside of Y-axis. Therefore, we need to measure whether the observed performance $d$ is in between $d^+$ and $d^-$ at each timestamp. The first two cases are straightforward. We define a threshold $Th$. When $Th$ percentage of the users' observed performance matches with case 1 or case 2, we either take no action or update the signature. The value of $T_h$ is set experimentally. In case 3, if $T_h$ percentage of users' performance is within the noise bandwidth and their shape does not match then we reduce the performance noise bandwidth. We experimentally define a similarity threshold $T_s$, which determines the minimum acceptable similarity between observed performance and the categorical signature. After reducing the bandwidth, we apply the change detection process again for each user's observed performance. In case 4, we increase the size of the noise bandwidth based on the observed performance and apply the change detection process again. We define a threshold $\delta d$, which determines how much noise bandwidth needs to be increased or decreased in cases 3 and 4. Value of $\delta d$ is set based on trials on the experiment.

\section{Experiment}

A series of experiments are conducted to evaluate the proposed change detection approach. We identify two key attributes: a) average delay and b) ability to detect changes or detection accuracy to evaluate the proposed approach. The proposed approach is compared with the existing IaaS performance changed detection approach proposed in \cite{fattah2020event}. 

\subsection{Experiment Setup}

The focus of this paper is to detect changes in IaaS performance behavior to keep the signature up to date. To evaluate the proposed framework's ability to detect changes, we require an environment where a set of consumers performs free trials on different services based on their workloads over different periods and observe service performance over a long period of time. We then require a scenario where service performance changes and impacts the experience of the trial users. Finding such real-world workload-performance dataset is challenging. To the best of our knowledge, there is no existing long-term workload-performance datasets of IaaS services available publicly. Therefore, we leverage existing short-term available datasets to synthesize datasets for our experiments. We use the Eucalyptus IaaS workload to generate the trial workloads of different consumers\footnote{\url{https://www.cs.ucsb.edu/~rich/workload/}}. It contains six workload traces of a production cloud environment. We select a trace that contains 34 days of workloads of a large company with 50,000 to 100,000 employees. We partition the data into 360 parts and consider each partition an average workload of day to create a 1-year workload data. The long-term performance of 5 IaaS providers is generated from the benchmark results published SPEC Cloud IaaS 2016 \cite{fattah2020icws}. We augment the workload traces with the performance data to generate a long-term workload-performance dataset of five IaaS providers. We create the signature of each provider using the approach in \ref{sec:sig}. The experiment variables are shown in Table \ref{tab:data}. We conduct the experiments by changing the signatures randomly to create new signatures. We have developed the experiment using Matlab on a computer with Intel Core i7 (2.80 GHz and 8Gb ram).  \textit{We have made our dataset and source code publicly available to make this experiment reproducible}\footnote{\url{https://github.com/sm-fattah/IaaS-Signature-Change-Detection-Experiment}}

%First, we map each unique workload of the trace to a unique performance value of the benchmark results. We consider the map as a baseline performance for the workload. Next, we build the long-term performance profiles for the providers where each provider shows different performance behavior based on the workloads and time. The behavior is generated using uniformly distributed random numbers. We run the workloads of each consumer on five providers to generate the performance data of each provider. 

\begin{table}[t]
%\vspace{-3mm}
\centering
\caption{Experiment Variables}\label{tab:data}
\begin{tabular}{|l|l|}
\hline
{\bfseries Variable Name} & {\bfseries Values}\\
\hline
Total provisioning period & {360}  days \\
Trial length of each consumer & {30} days \\
Total number of IaaS performance signatures & {5} \\
Total number of Consumers &  {18} \\
Similarity thresholds & { .6 to 0.9 }\\
Anomaly Thresholds & {60\% to 90\%} \\
\hline
\end{tabular}
\vspace{-5mm}
\end{table}

We identify the following two key variable in the experiment that drives the performance of the proposed approach:

\begin{itemize}
    \item Similarity Threshold: The similarity threshold indicates the minimum similarity between the shape of the observed performance in the trial of a consumer and the corresponding signature. The similarity threshold is utilized to determine shape-based similarity.
    \item Anomaly Threshold: The proposed change detection framework relies on the trial experience of the majority of the users. Based on the observation of the majority of the users, we either confirm change on update performance noise. The anomaly threshold defines the minimum number of users that are considered as the majority of the users. 
    
\end{itemize}

\subsection{Evaluation and Discussion}

\begin{figure}[!t]
%\vspace{-6mm}
    \centerline{
        \subfloat[]{\includegraphics[width=0.5\textwidth]{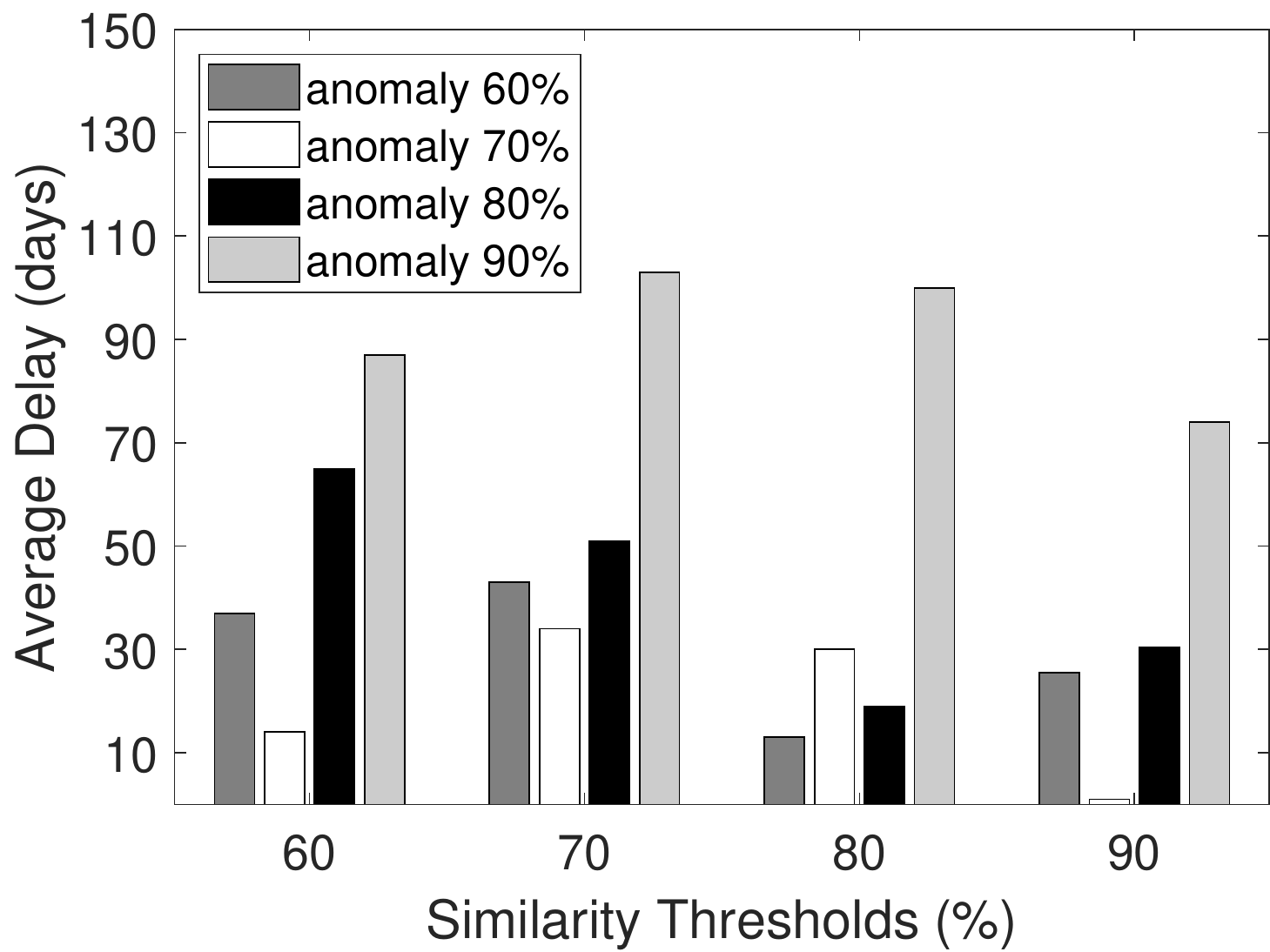}}
        \hfil
        \subfloat[]{\includegraphics[width=0.5\textwidth]{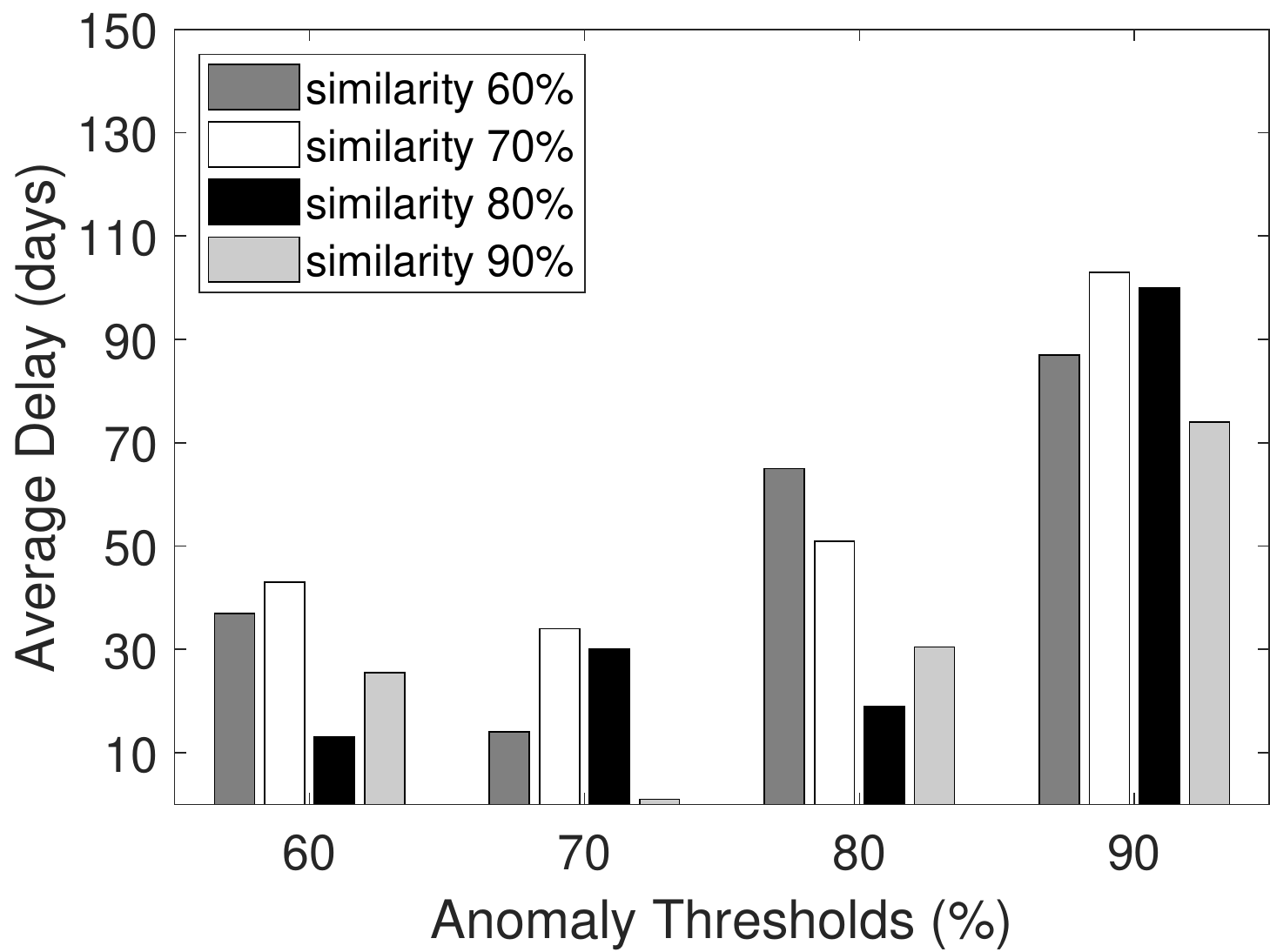}}
    }
     \centerline{
        \subfloat[]{\includegraphics[width=0.5\textwidth]{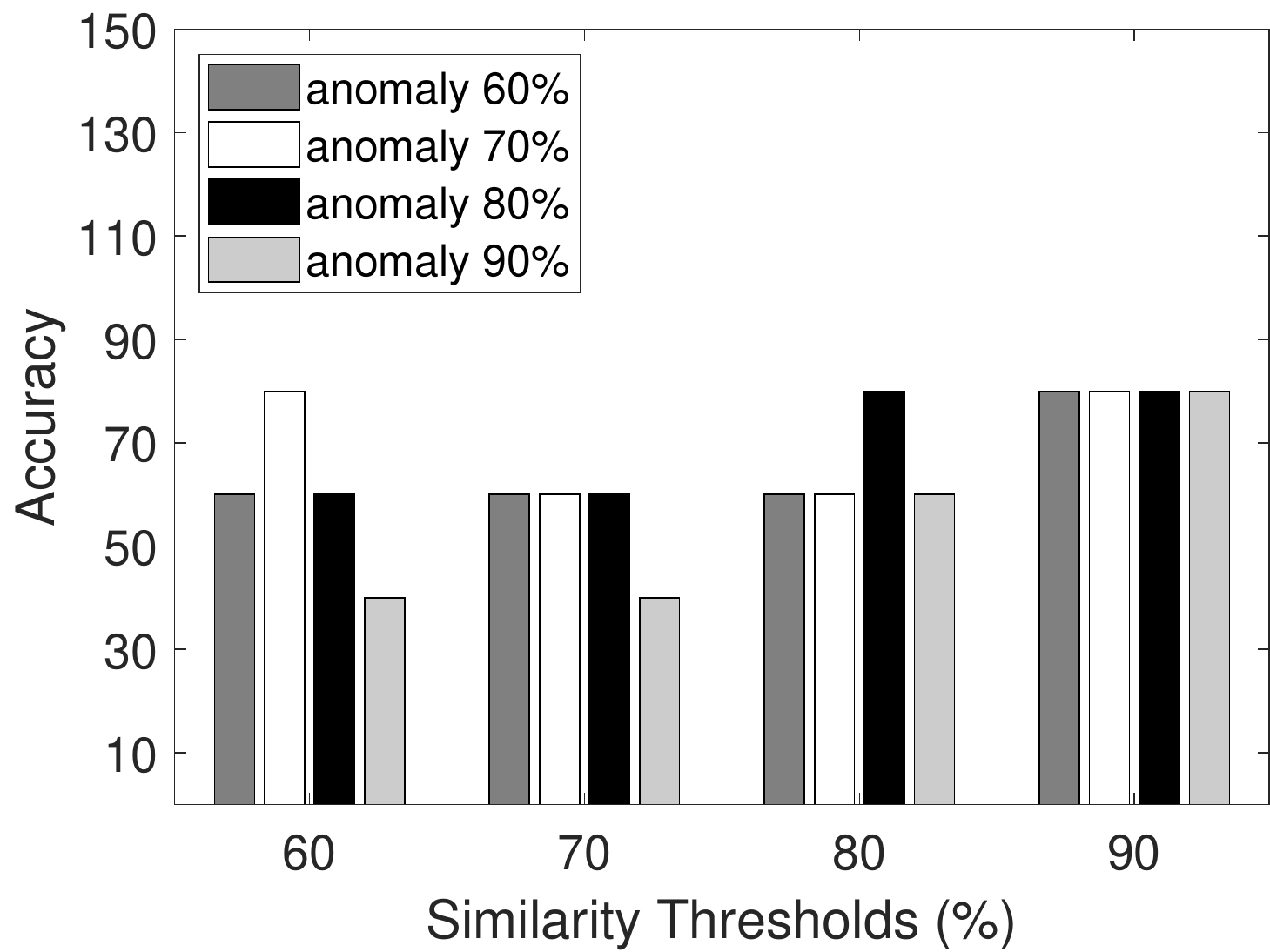}}
        \hfil
        \subfloat[]{\includegraphics[width=0.5\textwidth]{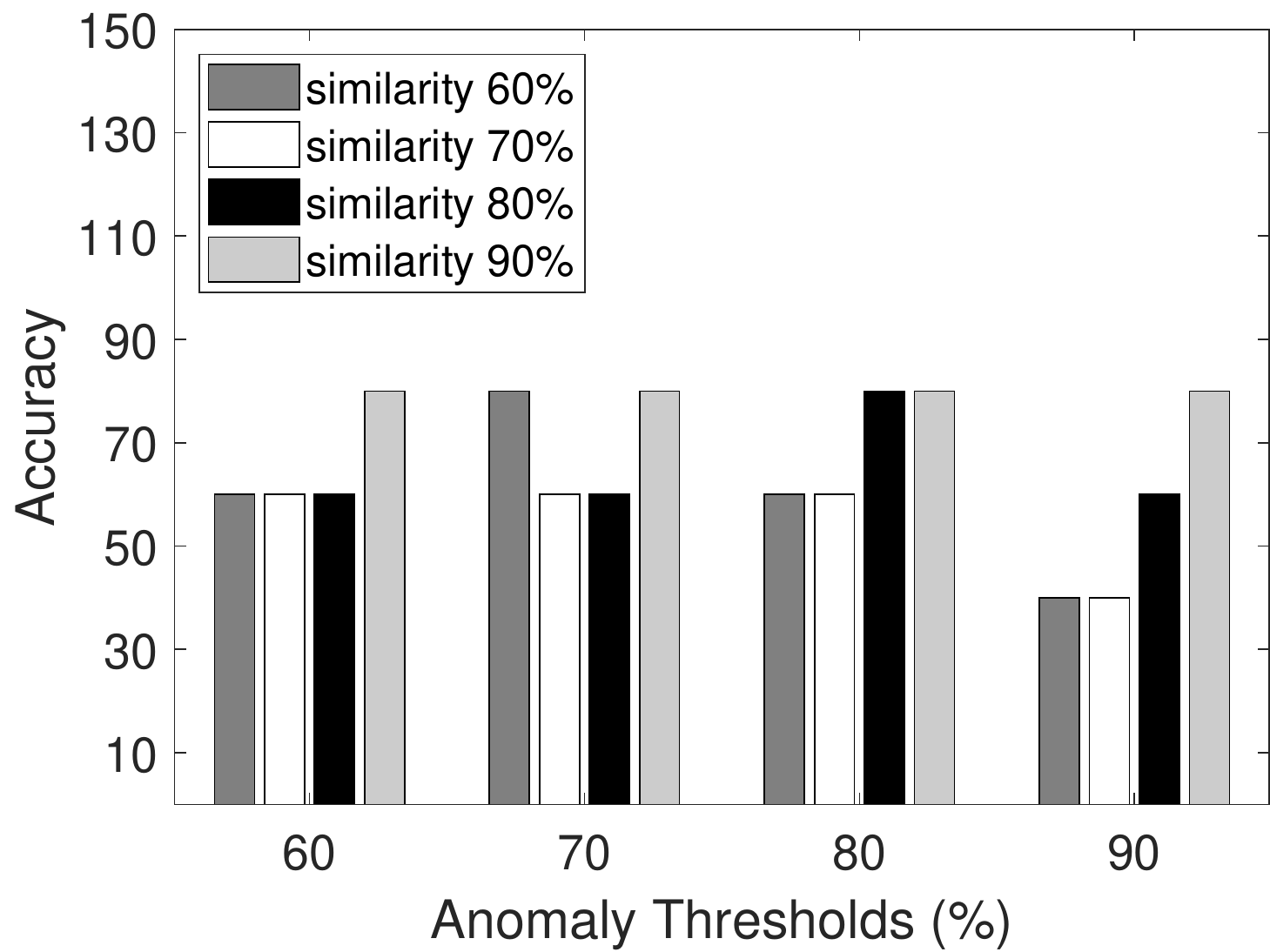}}
    }
 
    \caption{ (a) Average delay for variable similarity thresholds (b) Average delay for variable anomaly thresholds (c) Accuracy for variable similarity thresholds (d) Accuracy for variable anomaly thresholds}
  \label{fig:exp1}
  %\vspace{-6mm}
\end{figure}

We evaluate the proposed approach in terms of the average delay to detect signature changes and its ability to detect true changes in signature. The expectation is to reduce the average delay to detect the change in performance and increase the accuracy of detecting changes. Here, accuracy refers to the true positives, i.e., how many changes the proposed approach is able to detect. Fig. \ref{fig:exp1} depicts the results of experiments. Fig. \ref{fig:exp1}(a) and (b) show the average delay in detecting changes. Fig. \ref{fig:exp1}(a) shows the average delay for different similarity thresholds. There is no trend visible that indicates that there is a linear relationship between the similarity threshold and average change detection delay. The figure shows that the average delay is minimum when the similarity threshold is about 90\%. However, the average also depends on the anomaly threshold. When the anomaly threshold is about 70\%, the average delay is minimum in most cases in Fig. \ref{fig:exp1}(a). Similarly, Fig. \ref{fig:exp1}(b) shows the average delay for different anomaly thresholds. It also shows no common trend in the average detection delay based on the anomaly threshold. The average delay is minimum when the anomaly threshold is about 70\%, and the similarity threshold is about 80\%. 

The average delay is not the only attribute to measure the performance. We consider the accuracy of the proposed approach in terms of its ability to identify true changes correctly. Fig. \ref{fig:exp1}(c) and (d) show the accuracy of the proposed approach. In Fig. \ref{fig:exp1}(c), the accuracy is illustrated with respect to the different similarity thresholds. The accuracy of the proposed approach is about 80\% when the similarity threshold is 90\%. The effect of different anomaly thresholds is not very substantial on the accuracy according to the figure. Fig. \ref{fig:exp1}(d) illustrates the accuracy with respect to the anomaly threshold. When the anomaly threshold is about 90\%, that means 90\% of the users' experience does not match the corresponding signature, and the similarity threshold is about 90\%, the accuracy of the proposed approach is about 80\%. The proposed approach finds the changes in IaaS performance based on an iterative approach that conditionally updates the performance noise. Therefore, the change detection process stops when the suitable performance noise bandwidth is measured, confirming whether there is a change in the signature.

\begin{figure}[!t]
%\vspace{-4mm}
    \centerline{
        \subfloat[]{\includegraphics[width=0.5\textwidth]{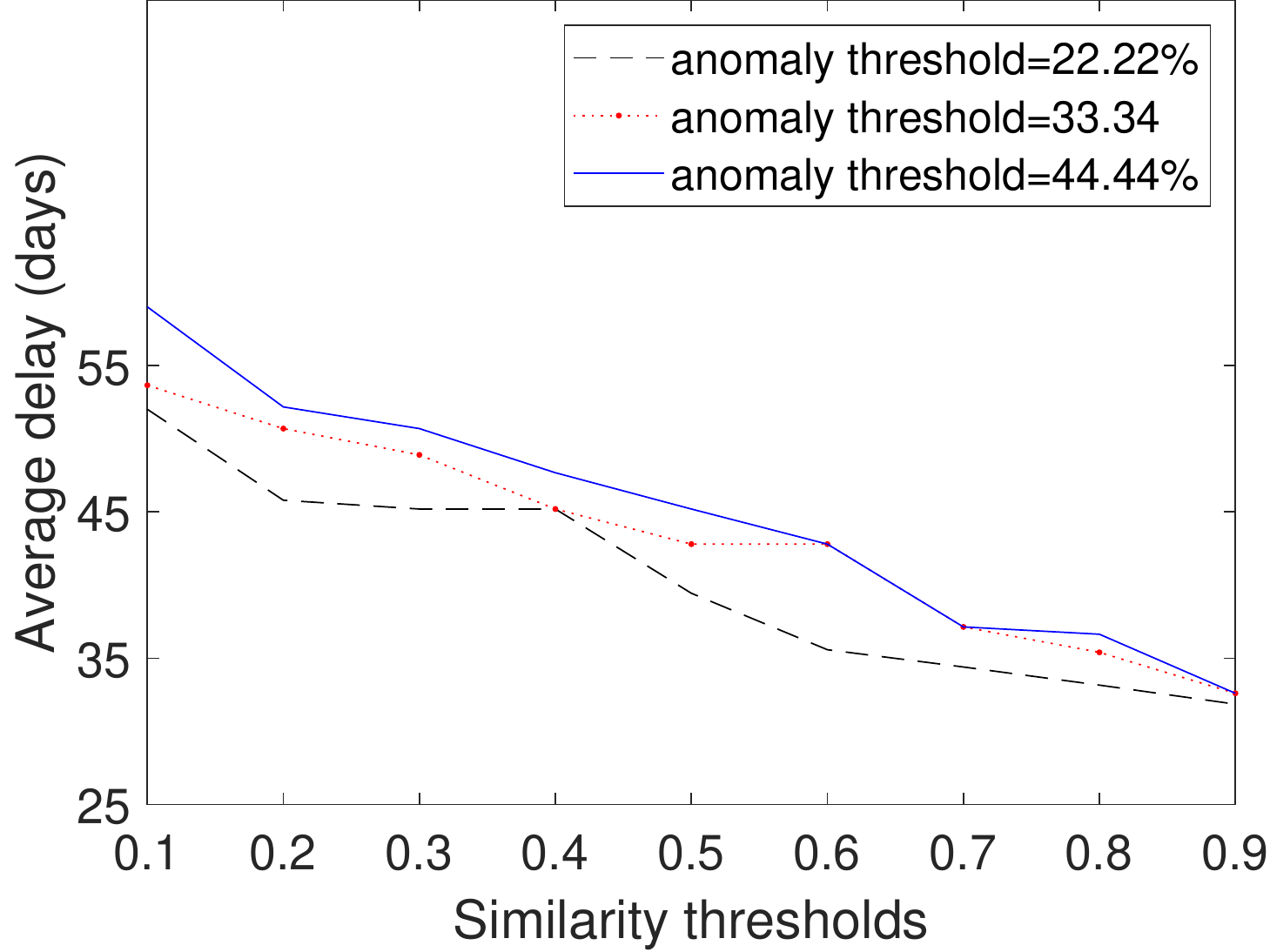}}
        \hfil
        \subfloat[]{\includegraphics[width=0.5\textwidth]{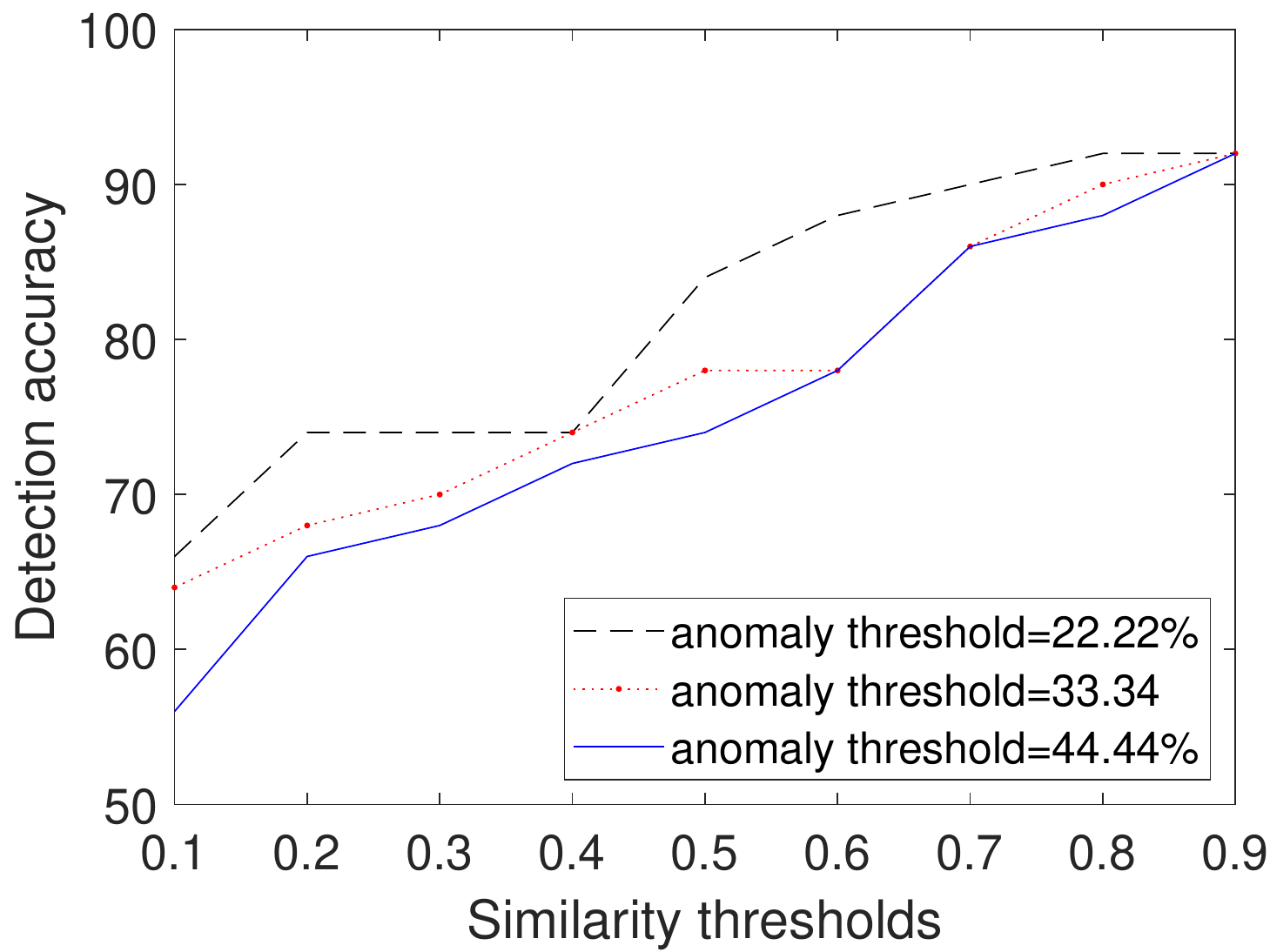}}
    }
 
    \caption{ Performance of the ECA approach (a) Average delay (b) Accuracy }
  \label{fig:exp2}
  \vspace{-6mm}
\end{figure}

\subsection{Comparison with Existing Work}

We have implemented the proposed ECA approach in \cite{fattah2020event} and applied it to our dataset. The result of the ECA approach is illustrated in Fig. \ref{fig:exp2}. Fig. \ref{fig:exp2}(a) shows the average delay for different similarity thresholds and anomaly thresholds in the ECA approach. The average delay in this approach can be 55 days to 35 days, depending on the similarity and anomaly thresholds. The average delay in our approach can be from 2 days to 110 days, depending on the similarity and the anomaly thresholds. Choosing the right similarity and anomaly threshold provides a better result than the ECA approach in terms of average change detection delay. The detection accuracy in Fig. \ref{fig:exp2} shows that the ECA approach provides accuracy from 60\% to 90\%, depending on the similarity and the anomaly threshold. The proposed approach in this work has an accuracy of about 60\% to 80\%. However, it does not produce any false positives where the proposed ECA approach in \cite{fattah2020event} produces a significant number of false positives.

\section{Related Work}

Performance is one of the most important criteria during cloud service selection \cite{iosup2011performance}. The performance of IaaS services has been studied in numerous studies \cite{iosup2014iaas,leitner2016patterns, wang2018testing, fattah2020icws}. An IaaS cloud service's performance is typically measured for different applications based on short-trials in IaaS cloud \cite{wang2018testing,fattah2020icws}. Most existing approaches do not consider the long-term performance variability of IaaS cloud services. IaaS performance has been extensively studied in \cite{leitner2016patterns}. The study suggests that cloud performance is a ``moving target" and requires re-evaluation periodically. A signature-based IaaS cloud service selection approach is proposed in \cite{fattah2020icws}. The proposed approach represents the long-term IaaS performance variability using the concept of the IaaS performance signature. The performance signature of an IaaS service is generated from the experience of the past trial users who share their data with a trusted third party. The trusted third party analyzes the periodic performance behavior of an IaaS service to generate its corresponding performance signature. However, the proposed work does not consider the changes in the signature over a long period of time or the effect of different types of workload in the performance of an IaaS service \cite{fattah2020icws}. 

To the best of our knowledge, there is no prior work that addresses the long-term IaaS performance change detection problem \cite{fattah2020icws}. The proposed approach in \cite{fattah2020event} mainly focuses on the change point detection (CPD) in IaaS performance. The CPD is a pre-requisite of IaaS performance change detection \cite{aminikhanghahi2017survey}. In the CPD problem, the distribution of data before and after the change is often considered known. The proposed work in \cite{fattah2020event} introduces an ECA model to detect change points in IaaS performance behavior. The ECA approach is an effective CPD technique. Other change point detection techniques include Bayesian change point detection, Shapelet, Model fitting, and Gaussian process. The work in \cite{fattah2020event} utilizes the CUSUM control chart to detect changes in IaaS performance. CUSUM relies on the mean and standard deviation of a time series to detect changes. However, CUSUM is unable to differentiate between noise and change in IaaS performance \cite{page1961cumulative}. Change detection in time series data is usually performed using different similarity measure techniques. However, most of these approaches do not consider the noise that may appear in the data. Therefore, we introduce a change detection framework that identifies noise in IaaS performance by leveraging the concept of categorical signature and noise bandwidth.

\section{Conclusion}

We propose a novel framework to detect long-term changes in IaaS performance behavior. The long-term performance behavior of an IaaS is represented by its performance signature. A new type of IaaS performance signature called categorical IaaS performance signature is introduced to capture the effect of different types of workload in the IaaS signature. The proposed framework introduces a signature change detection approach with performance noise. The key challenge in performance change detection is to differentiate between noise and accurate changes in IaaS performance. We introduce a new IaaS performance noise model to identify performance change accurately. The experiment results show that the proposed framework detects changes in IaaS performance effectively. We aim to investigate IaaS performance noise in more detail to develop more accurate change detection approaches in future work.

\section{Acknowledgement}

This research was partly made possible by DP160103595 and LE180100158 grants from the Australian Research Council. The statements made herein are solely the responsibility of the authors.

\bibliographystyle{splncs04}
\bibliography{references}

\end{document}